\documentclass[%
 aip,
 amsmath,amssymb,
 reprint,%
]{revtex4-1}

\usepackage{graphicx}
\usepackage{dcolumn}
\usepackage{bm}

\usepackage[utf8]{inputenc}
\usepackage[T1]{fontenc}
\usepackage{mathptmx}

\begin{document}


\title{Magnetization Plateau of the Distorted Diamond Spin Chain with Anisotropic Ferromagnetic Interaction}
\author{T\^oru Sakai}
\thanks{corresponding author}
\email[]{sakai@spring8.or.jp}
\altaffiliation{National Institutes for Quantum Science and Technology, SPring-8,
Hyogo 679-5148, Japan}
\author{Kiyomi Okamoto}%
\author{Hiroki Nakano}
\author{Rito Furuchi}
\affiliation{ 
Graduate School of Material Science, University of Hyogo, Hyogo 678-1279, Japan}

\date{\today}

\begin{abstract}
The $S=1/2$ distorted diamond spin chain with the anisotropic ferromagnetic interaction is investigated 
using the numerical diagonalization and the level spectroscopy analysis. 
It is known that the system exhibits a plateau of the magnetization curve at the 
1/3 of the saturation. The present study indicates that as the anisotropy is varied 
the quantum phase transition occurs between two different mechanisms of the 
1/3 magnetization plateau. The phase diagram with respect to the anisotropy 
and the ferromagnetic coupling is also presented. 

\end{abstract}


\maketitle

\def\vS{{\bf S}}  

\section{Introduction}

The magnetization plateau is one of interesting phenomena in the field 
of the magnetism. 
It possibly appears as the quantization of magnetization, 
when the one-dimensional quantum spin system satisfies the 
necessary condition
\begin{eqnarray}
   S_{\rm unit}-m_{\rm unit}={\rm integer},
\label{condition}
\end{eqnarray}
where $S_{\rm unit}$ is the total spin and $m_{\rm unit}$ is the magnetization per 
unit cell\cite{oshikawa}. 
The $S=1/2$ distorted diamond spin chain\cite{okamoto0} is a strongly frustrated 
quantum spin system which exhibits the 1/3 magnetization plateau 
according to the conditions. 
This system was proposed as a good theoretical model of 
the compound Cu$_3$(CO$_3$)$_2$(OH)$_2$, called azurite\cite{kikuchi,kikuchi2}.
In fact, the magnetization measurement of azurite 
detected a clear magnetization 
plateau at 1/3 of the saturation magnetization. 
The theoretical study by the numerical exact diagonalization\cite{okamoto1,okamoto2} 
suggested that the 1/3 magnetization plateau is induced by two different mechanisms. 
One is based on the ferrimagnetic mechanism and the other is due to the formation of 
singlet dimers at the $J_2$ bonds in Fig.1 and free spins. 
The 1/3 plateau of azurite is believed to be due to the latter mechanism \cite{kikuchi2,aimo}

Recently other candidate materials for the distorted diamond spin chain were discovered. 
They are the compound  K$_3$Cu$_3$AlO$_2$(SO$_4$)$_4$, called alumoklyuchevskite\cite{fujihara,morita,fujihala2}
and related materials.
All the exchange interactions of azurite are antiferromagnet, 
while alumoklyuchevskite includes ferromagnetic couplings as well as antiferromagnetic ones. 
The most important point is the the difference of the $J_2$ bond,
although the model of alumoklyuchevskite is more complicated than that of azurite.
The $J_2$ bond is antiferromagnetic on which a singlet dimer pair is formed in azurite,
whereas it is ferromagnetic in alumoklyuchevskite.
Thus it would be useful to investigate the distorted diamond chain with the 
ferromagnetic interaction. 
In this paper we investigate the distorted diamond spin chain model 
with the ferromagnetic exchange interaction and introduce the $XXZ$ 
anisotropy to this ferromagnetic bond. 
It is expected that this model would exhibit the 1/3 magnetization 
plateau based on two different mechanisms, as was predicted 
for the $(S, s)=(1, 1/2)$ mixed spin chain\cite{sakai-okamoto}. 
Using the numerical diagonalization of finite-size systems and 
the level spectroscopy analysis, we investigate the mechanisms of 
the 1/3 magnetization plateau in the present model and obtain 
the phase diagram with respect to the anisotropy and the 
strength of the ferromagnetic coupling.

\section{Model}

We investigate the model described by the Hamiltonian
\begin{eqnarray}
  {\cal H} &=& {\cal H}_0 + {\cal H}_{\rm Z} \\
  {\cal H}_0
   &=&  J_1 \sum_{j=1}^{N/3} \left[ \vS_{3j-2} \cdot \vS_{3j-1} 
          + \vS_{3j-1} \cdot \vS_{3j} \right] \nonumber \\
  &&+ J_2 \sum_{j=1}^{N/3} \left[ S_{3j}^x S_{3j+1}^x + S_{3j}^y S_{3j+1}^y 
    + \lambda S_{3j}^z S_{3j+1}^z \right]\nonumber \\
  ~~~~~~~~~~~ &&+ J_3 \sum_{j=1}^{N/3} \left[ \vS_{3j-1} \cdot 
  \vS_{3j+1}
          + \vS_{3j} \cdot \vS_{3j+2} \right] \\
  {\cal H}_{\rm Z}
   &=& -H \sum_{l=1}^{N} S_l^z 
   \label{eq:ham}
\end{eqnarray}

\begin{figure}[tbh]
\includegraphics[width=0.8\linewidth]{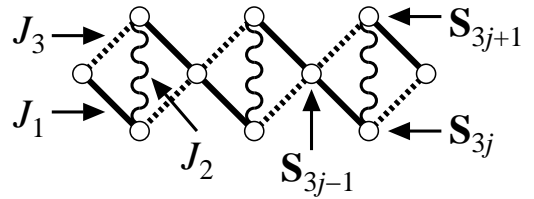}
\caption{The model of the $S=1/2$ distorted diamond spin chain. 
Solid lines, wavy lines and dotted lines denote $J_1$, $J_2$ and $J_3$, respectively.}
\label{model}
\end{figure}

\noindent
where $\vS_j$ is the spin-1/2 operator, $J_1$, $J_2$, $J_3$ are the 
coupling constants of the exchange interactions, respectively,
and $\lambda$ is the 
coupling anisotropy. 
The schematic picture of the model is shown in Fig. \ref{model}. 
In this paper we consider the case where $J_2$ is ferromagnetic 
and the $XY$-like (easy-plane) anisotropy is introduced to this bond only 
($J_2<0$ and $\lambda <1$), 
while $J_1$ and $J_3$ are isotropic antiferromagnetic bonds
($J_1, J_3 >0$ ). 
(We note that the case of $J_1<0$, $J_2>0$, $J_3 \gtrless 0$ and $\lambda =1$ was studied.
\cite{okamoto2005}).
$N$ is the number of spins and $L$ is defined as the number 
of the unit cells, namely $N=3L$. 
For $L$-unit systems, 
the lowest energy of ${\cal H}_0$ in the subspace where 
$\sum _j S_j^z=M$, is denoted as $E(L,M)$. 
The reduced magnetization $m$ is defined as $m=M/M_{\rm s}$, 
where $M_{\rm s}$ denotes the saturation of the magnetization, 
namely $M_{\rm s}=3L/2$ for this system. 
$E(L,M)$ is calculated by the Lanczos algorithm under the 
periodic boundary condition ($ {\bf S}_{N+l}={\bf S}_l$) 
and the twisted boundary condition 
($S^{x,y}_{N+l}=-S^{x.y}_l, S^z_{N+l}=S^z_l$), 
for $L=$4, 6 and 8. 
Under the twisted boundary condition 
we calculate the lowest energy $E_{{\rm TBC},P=+}(L,M)$
($E_{{\rm TBC},P=-}(L,M)$) 
in the subspace where 
the parity is even (odd) with respect to the lattice inversion 
$\vS_l \to \vS_{N-l+1}$ at the twisted bond.

\section{Magnetization plateau}

\begin{figure}[tbh]
\includegraphics[width=0.8\linewidth]{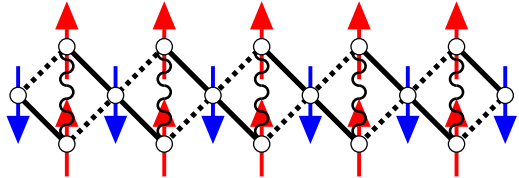}
\caption{Schematic picture of the ferrimagnetic mechanism 
of the 1/3 magnetization plateau. 
}
\label{ferri-1}
\end{figure}

\begin{figure}[tbh]
\includegraphics[width=0.8\linewidth]{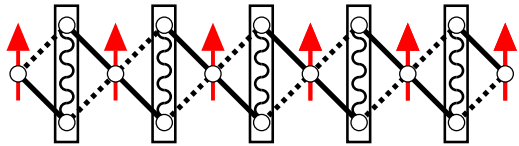}
\caption{Schematic picture of another mechanism 
of the 1/3 magnetization plateau, induced by the 
strong easy-plane anisotropy of the ferromagnetic $J_2$ bond.
Here the rectangles denote the two spin state $(1/\sqrt{2})(\uparrow\downarrow + \downarrow \uparrow)$.
}
\label{ferri-2}
\end{figure}

In the isotropic coupling case $\lambda=1$ 
the model (2) is the ferrimagnet with the 
1/3 spontaneous magnetization and has the
1/3 magnetization plateau. 
The previous numerical diagonalization and the 
level spectroscopy study\cite{sakai-okamoto} 
on the $(S,s)=(1,1/2)$ 
mixed spin ferrimagnetic chain indicated that 
as the easy-plane anisotropy increases, 
the system exhibits the quantum phase transition 
at the critical point where the ferrimagnetic 
magnetization plateau disappears and a new 
plateau due to another mechanism appears. 
The equivalent quantum phase transition is 
expected to occur in the present model. 
The initial 1/3 magnetization plateau 
is based on the ferrimagnetic mechanism 
shown in Fig. \ref{ferri-1}, 
while another mechanism of the 1/3 plateau 
is expected to be due to a strong easy-plane 
anisotropy effect shown in Fig. \ref{ferri-2}. 
We denote these two plateaux as the plateaux I 
and II, respectively.

\begin{figure}
\includegraphics[width=0.85\linewidth,angle=0]{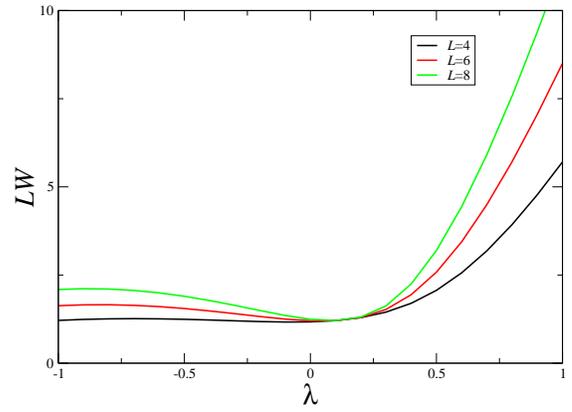}%
\caption{\label{prg}
Scaled plateau width $LW$ for $L=$4, 6 and 8 is plotted versus 
the anisotropy $\lambda$ in the case of $J_1=1.0, J_2=-1.0$ and $J_3=0.5$
 }
\end{figure}

In order to investigate the quantum phase transition 
between the plateaux I and II, 
we apply the phenomenological renormalization 
for the plateau width $W=E(L,M+1)+E(L,M-1)-2E(L,M)$ 
at $M=L/2$ calculated by the numerical diagonalization. 
The scaled plateau width $LW$ for $L=$4, 6 and 8 is plotted versus 
the anisotropy $\lambda$ in the case of $J_1=1.0, J_2=-1.0$ and $J_3=0.5$ 
shown in Fig. \ref{prg}. 
It indicates the quantum phase transition around $\lambda \sim 0.2$ 
where the first plateau vanishes 
and the other plateau appears with decreasing $\lambda$.  
The phase boundary will be estimated in the next section. 

\section{Level spectroscopy and phase diagram}

In order to detect the quantum phase transitions among the 
plateaux I, II and plateauless phases, 
the level spectroscopy analysis\cite{kitazawa1,kitazawa2} is one of the best methods. 
According to this analysis, 
we should compare the following three energy gaps; 
\begin{eqnarray}
\label{delta2}
&&\Delta _2 ={E(L,M-2)+E(L,M+2)-2E(L,M) \over 2}, \\
\label{tbc+}
&&\Delta_{{\rm TBC},P=+}=E_{{\rm TBC},P=+}(L,M)-E(L,M), \\
\label{tbc-}
&&\Delta_{{\rm TBC},P=-}=E_{{\rm TBC},P=-}(L,M)-E(L,M).
\end{eqnarray}

\begin{figure}[h]
\bigskip
\bigskip
\includegraphics[width=0.85\linewidth,angle=0]{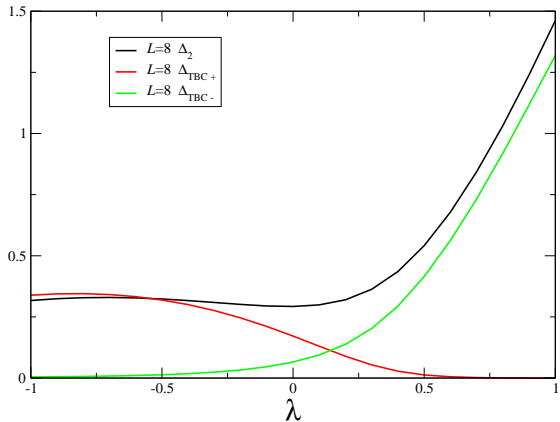}%
\caption{\label{LS}
Level spectroscopy analysis applied to the present model for 
 $J_1=1.0$, $J_2=-1.0$ and $J_3=0.5$. 
 }
\bigskip
\end{figure}

The level spectroscopy method indicates that the smallest gap 
among these three gaps for $M=L=M_{\rm s}/3$ determines the phase 
at $m=1/3$. 
$\Delta_2$, $\Delta_{{\rm TBC},P=+}$ and $\Delta _{{\rm TBC},P=-}$ 
correspond to the plateauless, plateau I and plateau II phases, 
respectively. 
Especially, $\Delta_{{\rm TBC,}P=+}$ and  $\Delta_{{\rm TBC},P=-}$ directly reflect the symmetries of two plateau states.
The physical explanation is very simple.
In the ferrimagnetic mechanism,
the state is essentially composed of the direct product of local state,
which leads to no change ($P=+$) under the space inversion operation.
On the other hand,
there is a two-spin state $(1/\sqrt{2})(\uparrow_N \downarrow_1 + \downarrow_1 \uparrow_N)$ on the $J_2$ bonds
connecting $\vS_N$ and $\vS_1$.
This state changes into the $(1/\sqrt{2})(\uparrow_N \downarrow_1 - \downarrow_1 \uparrow_N)$ state
if the corresponding $J_2$ bond is twisted.
When we perform the space inversion operation,
this state becomes $(1/\sqrt{2})(\downarrow_1 \uparrow_N - \uparrow_1 \downarrow_N)$,
which means $P=-$.
The $\lambda$ dependence of these three gaps for $J_1=1.0$, $J_2=-1.0$ 
and $J_3=0.5$ is shown in Fig. \ref{LS} for $L=$8. 
It indicates that the quantum phase transition between the plateaux I and II phases occurs, 
but the plateauless phase does not appear.

\begin{figure}
\includegraphics[width=0.85\linewidth,angle=0]{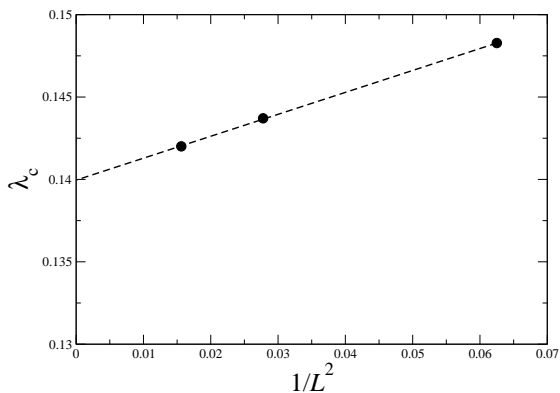}%
\caption{\label{extrap}
Process of the extrapolation of the phase boundary to the thermodynamic limit $L\rightarrow \infty$. 
The model parameters are the same as those of Fig.\ref{LS}. }
\end{figure}

Assuming that the finite-size correction is proportional to $1/L^2$, 
we estimate the phase boundaries in the thermodynamic limit 
from every level-cross point. 
The process for the same parameters as Fig. \ref{LS} is shown in Fig. \ref{extrap}. 

\begin{figure}
\bigskip
\bigskip
\bigskip
\includegraphics[width=0.85\linewidth,angle=0]{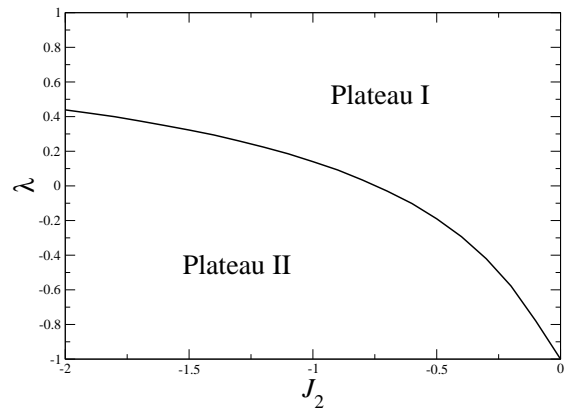}%
\caption{\label{phase}
Phase diagram on the $J_2$-$\lambda$ plane at $m=1/3$ 
for $J_1=1.0$ and $J_3=0.5$. 
 }
\end{figure}

The phase diagram on the $J_2$-$\lambda$ plane at $m=1/3$ 
for $J_1=1.0$ and $J_3=0.5$ 
is shown in Fig. \ref{phase}. 
The two different plateau phases appears, but the plateauless phase does not appear. 
The plateau II phase is predicted for the first time at least for the 
distorted diamond spin system.

\section{Summary}

The $S=1/2$ distorted diamond spin chain with the anisotropic ferromagnetic interaction is investigated 
using the numerical diagonalization and the level spectroscopy analysis. 
As a result it is found that as the easy-plane anisotropy of the ferromagnetic bond increases 
the quantum phase transition occurs at the critical point where the mechanism of the 1/3 
magnetization plateau changes. 
The phase diagram with respect to the strength of the ferromagnetic coupling and the anisotropy 
is also presented.

\begin{acknowledgments}
This work was partly supported by JSPS KAKENHI, 
Grant Numbers JP16K05419, JP20K03866, JP16H01080 (J-Physics), 
JP18H04330 (J-Physics) and JP20H05274.
A part of the computations was performed using
facilities of the Supercomputer Center,
Institute for Solid State Physics, University of Tokyo,
and the Computer Room, Yukawa Institute for Theoretical Physics,
Kyoto University.
\end{acknowledgments}

\section*{Data Availability}
The data that support the findings of this study are available from the corresponding author upon reasonable request.

\nocite{*}
\bibliography{aipsamp}

\end{document}